    \documentclass{JHEP3}
\usepackage{amsmath}
\usepackage{amssymb}
\usepackage{graphicx}
\numberwithin{equation}{section}        

\usepackage{epsfig}

\numberwithin{equation}{section}

    \renewcommand{\(}{\left(}
    \renewcommand{\)}{\right)}

    \newcommand{\ph}{{\rm ph}}
    \newcommand{\mir}{{\rm mir}}
    \newcommand{\cut}{{\circlearrowright\hspace{-3.4mm}*\hspace{0.2mm}}}
    
    \newcommand{\YF}{Y_{\fF}}
    \newcommand{\Yf}{Y_{\ff}}
    \newcommand{\Ym}[1]{Y_{\fm_#1}}
    
    \newcommand{\Yb}[1]{Y_{\fb_#1}}
    \newcommand{\Yp}[1]{Y_{\fp_#1}}

    \newcommand{\ff}{\scalebox{0.67}{\begin{picture}(5.21,0.0)\put(-1.5,1.2){$\oplus$}\end{picture}}}
    \newcommand{\fF}{\scalebox{0.67}{\begin{picture}(5.21,0.0)\put(-1.5,1.2){$\otimes$}\end{picture}}}
    \newcommand{\fm}{\scalebox{1.15}{\begin{picture}(2.8,0.0)\put(-1.2,-0.33){$\bullet$}\end{picture}}}
    
    \newcommand{\fb}{\scalebox{0.5}{\begin{picture}(7.27,0.3)\put(-1.7,2.65){$\bigcirc$}\end{picture}}}
    \newcommand{\fp}{\scalebox{0.6}[0.7]{\begin{picture}(7.24,0.2)\put(-1.5,0.5){$\bigtriangleup$}\end{picture}}}

    \newcommand{\beq}{\begin{equation}}
    \newcommand{\eeq}{\end{equation}}
    \newcommand\beqa{\begin{eqnarray}}
    \newcommand\eeqa{\end{eqnarray}}
    \newcommand\bea{\begin{array}}
    \newcommand\eea{\end{array}}

    \newcommand{\nn}{\nonumber}

    \newcommand{\COMMENT}[1]{}
    
    \newcommand{\neqa}{\nonumber\end{eqnarray}}
    \newcommand{\la}[1]{\label{#1}}

    \newcommand{\eq}[1]{(\ref{#1})}

    \def\tr{{\rm tr~}}

    \renewcommand{\d}{\partial}

    \newcommand{\re}{\relax{\rm I\kern-.18em R}}

    \newcommand{\rb}{\right)}
    \newcommand{\lb}{\left(}

    \def\su2{{SU(2)}}

    \def\i2{\frac{i}{2}}

    \def\const{\mbox{const}}

    \newcommand{\figf}{\scalebox{0.67}{\begin{picture}(5.21,0.0)\put(-1.5,1.2){$\oplus$}\end{picture}}}
    \newcommand{\figF}{\scalebox{0.67}{\begin{picture}(5.21,0.0)\put(-1.5,1.2){$\otimes$}\end{picture}}}
    \newcommand{\figm}{\scalebox{1.15}{\begin{picture}(2.8,0.0)\put(-1.2,-0.33){$\bullet$}\end{picture}}}
    
    \newcommand{\figb}{\scalebox{0.5}{\begin{picture}(7.27,0.3)\put(-1.7,2.65){$\bigcirc$}\end{picture}}}
    \newcommand{\figp}{\scalebox{0.6}[0.7]{\begin{picture}(7.24,0.2)\put(-1.5,0.5){$\bigtriangleup$}\end{picture}}}

    \newcommand{\fo}{\scalebox{0.67}{\begin{picture}(5.21,0.0)\put(-1.5,1.2){$\blacktriangleleft$}\end{picture}}}
    \newcommand{\fO}{\scalebox{0.67}{\begin{picture}(5.21,0.0)\put(-1.5,1.2){$\blacktriangleright$}\end{picture}}}

    \newcommand{\be}{\begin{equation}}
    \newcommand{\ee}{\end{equation}}
    \newcommand{\bg}{\begin{gather}}
    \newcommand{\eg}{\end{gather}}
    \newcommand{\bseq}{\begin{subequations}}
    \newcommand{\eseq}{\end{subequations}}

    \newcommand{\cutclock}{{\circlearrowright\hspace{-3.48mm}*\hspace{1mm}}}

        \renewcommand{\Im}{{\rm Im}}
        \renewcommand{\Re}{{\rm Re}}

    \title{Numerical results for the exact spectrum of planar AdS4/CFT3}

        \author{Fedor~Levkovich-Maslyuk\\
        Blackett Laboratory, Imperial College, London SW7 2AZ, U.K. \&\\
        Institute of Theoretical and Experimental Physics,\\
        B. Cheremushkinskaya ul. 25, 117259 Moscow, Russia \\
 \email{fedor.levkovich$\bullet$gmail.com}}

    \abstract{
    We compute the anomalous dimension for a short single-trace operator in planar ABJM theory at intermediate coupling. This is done by solving numerically the set of Thermodynamic Bethe Ansatz equations which are expected to describe the exact spectrum of the theory. We implement a truncation method which significantly reduces the number of integral equations to be solved and improves numerical efficiency. Results are obtained for a range of 't Hooft coupling $\lambda$ corresponding to $0 \leq h(\lambda) \leq 1$, where $h(\lambda)$ is the interpolating function of the AdS$_4$/CFT$_3$ Bethe equations. 
    }
    \preprint{Imperial/TP/11/FLM/01\\ITEP-TH-11/12}

\begin{document}

    %%%%%%%%%%%%%%%%%%%%%%%%%%%%%%%%%%%%%
    % Begin main text
    %%%%%%%%%%%%%%%%%%%%%%%%%%%%%%%%%%%%%
    %%%%%%%%%%%%%%%%%%%%%%%%%%%%%%%%%%%%%

\section{Introduction}
Recently there has been a significant interest towards integrable structures which arise in the context of AdS/CFT correspondence \cite{Beisert:2010jr}, with the best-studied example being the AdS$_5$/CFT$_4$ duality between four-dimensional planar ${\cal N}=4$ Super--Yang-Mills theory and Type IIB superstring theory on ${{\rm AdS}_5\times {\rm S}^5}$. There are also other AdS/CFT dual pairs \cite{adscftintegr, Klose:2010ki} where integrability gives us serious hopes of solving exactly a highly non-trivial quantum field theory. One of these is the ABJM duality proposed in \cite{Aharony:2008ug}, which relates three-dimensional planar ${\cal N}=6$ super Chern-Simons theory and Type IIA string theory on ${\rm AdS}_4\times {\mathbb{CP}}^3$. It appears that the spectrum of anomalous dimensions/string state energies in this AdS$_4$/CFT$_3$ duality may be found exactly at any value of the coupling  \cite{Klose:2010ki} by applying an approach similar to the one used in AdS$_5$/CFT$_4$. In particular, the Bethe ansatz equations, which describe the spectrum for asymptotically long single-trace operators at any coupling, have been proposed in \cite{Minahan:2008hf} and extended to all loops in \cite{Gabacp3}, \cite{AhnBae}. These equations however do not capture the so-called wrapping interactions \cite{ambjorn} which means that other tools have to be used in order to obtain the spectrum for short operators or the energies in finite volume.

In the AdS$_5$/CFT$_4$ case this issue has been successfully addressed by means of the generalized Luscher formulae \cite{luscherref} and the Y-system/Thermodynamic Bethe ansatz (TBA) approaches \cite{GKV}, \cite{tbarefsads5}. For AdS$_4$/CFT$_3$ a Y-system of functional equations was proposed in \cite{GKV}, and was later refined as well as supplemented with a set of TBA integral equations in \cite{Bombardelli:2009xz}, \cite{GFcp3}. Unlike the Bethe ansatz, the TBA and Y-system are expected to be valid for any state, thus providing a way to obtain, in principle, the full exact spectrum of the theory.

For AdS$_5$/CFT$_4$ the Y-system and TBA have passed a number of nontrivial tests. In particular, the Y-system allows one to efficiently reproduce perturbative gauge theory calculations at weak coupling (see e.g. \cite{GKV}), while at strong coupling it matches the semiclassical predictions obtained from the algebraic curve \cite{ads5quasicl},  \cite{Gromov:2011de}. In addition to these analytical checks, important numerical results have been obtained from the TBA in \cite{GKVKonishi}, \cite{Frolov:2010wt} where the anomalous dimension of the Konishi operator was computed for a wide range of values of the 't Hooft coupling. The strong-coupling predictions obtained in these works seem to agree with the analytical results obtained by several other methods \cite{KonOther}, \cite{Gromov:2011de} (see also \cite{BassoKon}), providing yet another successful test of the proposed TBA and Y-system. Also, very recently the TBA equations were reduced to a finite set of integral equations \cite{Gromov:2011cx}.

Analogous checks have been done for AdS$_4$/CFT$_3$ -- the four-loop wrapping corrections obtained in \cite{GKV} were reproduced in \cite{MSS1}, \cite{MSS2}, while in \cite{GFcp3} the proposed Y-system and TBA were shown to be remarkably consistent with the algebraic curve quantization \cite{Gromov:2008bz}. However, a numerical analysis similar to \cite{GKVKonishi}, \cite{Frolov:2010wt} has not been attempted so far and would be an important test of the integrability properties in this theory.

Here we present a first step in this direction, solving numerically the TBA equations for one of the simplest unprotected operators in the $sl(2)$ sector to compute its anomalous dimension non-perturbatively. We start from weak coupling and are able to reach those values of the 't Hooft coupling $\lambda$ for which the AdS$_4$/CFT$_3$ interpolating function $h(\lambda)$ becomes equal to 1 (this should correspond to $\lambda \sim 1$ as well) \footnote{Since the TBA equations include $h(\lambda)$ rather than $\lambda$, our result is the anomalous dimension as a function of $h(\lambda)$ and not of $\lambda$.}. To facilitate efficient numerical analysis, we implement a truncation method \cite{Gtrunc} for the TBA equations which is based on partially solving the underlying Y-system and allows us to eliminate from the equations an infinite number of unknown functions.

%This should correspond to the intermediate coupling regime, $\lambda \sim 1$. We find that for $h(\lambda)\approx1$ one of the Y-functions gets very close to the critical value $-1$ which makes it a challenge to proceed further.

This paper is organized as follows. In section 2 we introduce the state that we are studying, in section 3 we present the TBA equations, and in section 4 describe the truncation method. In section 5 we discuss the numerical results, and we conclude in section 6. Appendix A contains notation that we use.

\section{Description of the state}
The gauge theory operator that we study in this paper is the $L=2$ state in irrep ${\bf 20}$ of $SU(4)$, described in the Appendix of \cite{Minahan:2008hf}. Grouping the scalar fields of the theory into $SU(4)$ multiplets as follows:
\be
	Y^A=(A_1,A_2,B^\dag_{\dot1},B^\dag_{\dot2})\qquad
	Y^\dag_A=(A^\dag_1,A^\dag_2,B_{\dot1},B_{\dot 2})\,,
\ee
we can write this operator as
\beq
\label{su2oper}
	{\cal O}=\tr(Y^{C}Y^\dag_{A}Y^{D}Y^\dag_{B})\;\chi^{AB}_{CD}\, ,
\eeq
where the coefficients $\chi^{AB}_{CD}$ are antisymmetric in both $A,B$ and $C,D$ pairs of indices (for more details see \cite{Minahan:2008hf}). This is one of the simplest unprotected operators in the ${\cal N}=6$ supersymmetric Chern-Simons theory. In the asymptotic Bethe ansatz (ABA) of \cite{Gabacp3} this state is described in $su(2)$ grading by two momentum-carrying Bethe roots -- one $u_4$ and one $u_{\bar 4}$, which are equal. The Bethe equations in this case reduce to\footnote{the dressing factor $\sigma$ does not appear because $\sigma(u,u)=1$.}
\beq
\label{su2bae}
        \(\frac{x_4^+}{x_4^-}\)^L=1
\eeq
where the function $x(u)$ is defined as
\be
    \label{xdef}
        x + \frac{1}{x} = \frac{u}{h(\lambda)},
    \ee
with standard two branches    
    \beq
    \label{xBranches}
        x^{\ph}(u)=\frac{1}{2}\lb
        \frac{u}{h}+\sqrt{\frac{u}{h}-2}\;\sqrt{\frac{u}{h}+2}\rb\;\;,\;\;
        x^{\mir}(u)=\frac{1}{2}\lb \frac{u}{h}+i\sqrt{4-\frac{u^2}{h^2}}\rb \, ,
    \eeq
and in the Bethe ansatz equations we use the physical branch. We also used the general notation
\be
        f^\pm \equiv f(u\pm i/2),\  f^{[+a]} \equiv f(u + ia/2)\ .
\ee
The function $h(\lambda)$ in \eq{xdef} is the so-called interpolating function (see \cite{Grig1}) which plays the role of the effective coupling in the Bethe ansatz and TBA. Its weak and strong coupling expansions are known to be
\beq
    h(\lambda)=\lambda+h_3\lambda^3+{\cal O}\(\lambda^5\)=
    \sqrt{\lambda/2}+h^0+{\cal O}\(\frac{1}{\sqrt{\lambda}}\)\;.
\eeq
The weak-coupling coefficient $h_3=-8+2\zeta(2)$ was computed directly from perturbation theory in \cite{MSS1}, \cite{MSS2}, \cite{Leoni:2010tb}. For the strong-coupling coefficient $h^0$ several calculations suggest different values: \cite{Sh}, \cite{Bergman:2009zh}, \cite{Abbott:2010yb}, \cite{Astolfi:2011ju}, \cite{Astolfi:2011bg} argue that it is zero, while \cite{McLoughlin:2008he} propose the value $-\frac{\log2}{2\pi}$ (see also \cite{Bombardelli:2008qd})\footnote{Related calculations at strong coupling on the string side of the duality have been done in \cite{GrigFinSize}.}.
    
The TBA equations correspond to the $sl(2)$ rather than $su(2)$ Bethe ansatz equations, so we need to make a duality transformation \cite{Gabacp3} in the Bethe ansatz. We find that the $sl(2)$ representative of this state is described by the same two Bethe roots but has $L=1$ rather than $L=2$, the corresponding Bethe equation being
\beq
\label{sl2bae}
        \(\frac{x_4^+}{x_4^-}\)^L=-1\ .
\eeq
The corresponding single-trace operator is in the same supermultiplet as \eq{su2oper} (and has the same anomalous dimension). Its explicit form can be found in \cite{Zwiebel:2009vb} \footnote{Roughly speaking, it is a linear combination of terms with two scalar fields and two covariant derivatives} and its bare dimension is 3.

Importantly, from the ABA equations \eq{sl2bae} we see that for any coupling the two roots remain exactly at zero\footnote{This is also consistent with the zero-momentum constraint \cite{Gabacp3}:
$\prod\limits_j\frac{x_{4,j}^+}{x_{4,j}^-}\prod\limits_j\frac{x_{\bar4,j}^+}{x_{\bar4,j}^-}$=1.}:
$u_4=u_{\bar4}=0$. This means that, unlike what happens for Konishi in the ${\cal N}=4$ SYM case, the Bethe ansatz result for the scaling dimension can be found exactly (in terms of the interpolating function $h(\lambda)$). It is given by
\beq
        E_{ABA}=E_0+\epsilon_1^{\ph} (u_4)+\epsilon_1^{\ph} (u_{\bar4})
\eeq
where $E_0$ is the bare dimension of the operator and 
\beq\label{epsilon}
        \epsilon_n (u) \equiv
        h(\lambda) \left( \frac{i}{x^{[+n]}} - \frac{i}{x^{[-n]}}\right),
\eeq
which means that in our case
\beq
\label{eaba}
        E_{ABA}=\sqrt{16h\(\lambda\)^2+1}+2 \ .
\eeq

It is well-known that the ABA result is usually incomplete because of the wrapping interactions which arise due to the finite length of the corresponding spin-chain \cite{ambjorn}. The {\it exact} anomalous dimension includes a correction, $\delta E$, to the ABA result:
\beq
\label{eplusde}
        E=E_{ABA}+\delta E\ .
\eeq
The leading weak-coupling term of this correction (first wrapping) has been computed in \cite{GKV} and confirmed in \cite{MSS2} (see also \cite{MGFwrap} where wrapping corrections for similar operators were studied). It has the form\footnote{since at weak coupling $h(\lambda)=\lambda+O(\lambda^2)$ this expression can be rewritten also as $\delta E= (32-16\zeta(2)) \lambda^4 + O(\lambda^5)$}
\beq
\label{1wrap}
        \delta E_{wrapping}= (32-16\zeta(2)) h(\lambda)^4 + O(\lambda^5)
\eeq
and thus
\beq
\label{aba1stwrap}
        E_{ABA}+ \delta E_{wrapping}=3+8h(\lambda)^2-16\zeta(2)h(\lambda)^4+O(\lambda^5).
\eeq
Our main result in this work is a numerical computation of the correction $\delta E$ for $0\leq h(\lambda)\leq 1$ by means of the Thermodynamic Bethe Ansatz.

\section{TBA equations for AdS4/CFT3}   

The Y-system which describes the spectrum of the ABJM theory in the planar limit was first proposed in \cite{GKV} and later refined in \cite{GFcp3}, \cite{Bombardelli:2009xz}. This system of functional equations can be summarized in the diagram shown in figure \ref{Fig:figabjmysys}.

\FIGURE[ht]
{\label{Fig:figabjmysys}

    \begin{tabular}{cc}
    \includegraphics[scale=0.6]{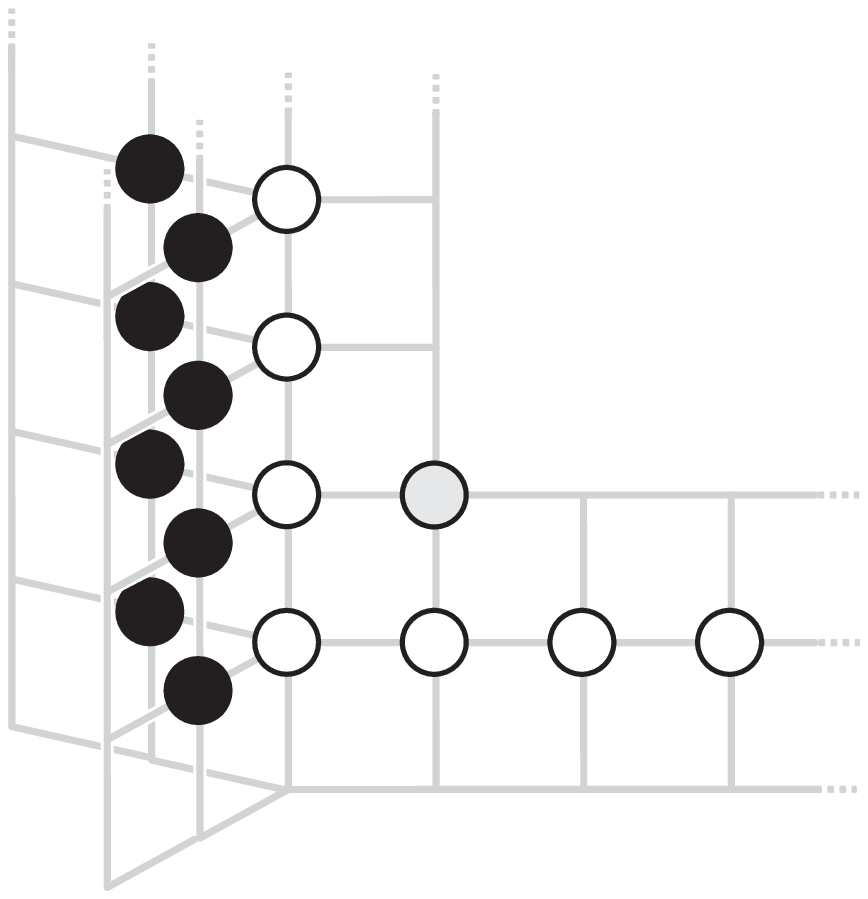}
    \end{tabular}
    \caption{A graphical representation of the Y-system proposed in \cite{GKV} for ABJM theory.
    Each circle in this infinite 3D lattice corresponds to a Y-function.}
}

We denote the various Y-functions as in \cite{GFcp3} -- there are fermionic functions $\Yf$ and $\YF$, bosonic functions $\Yb{n}$ ($n=2,3,\dots$), pyramid functions $\Yp{n}$ ($n=2,3,\dots$) and middle node functions $Y_{{\fO}_{n}}, Y_{{\fo}_{n}}$ ($n=1,2,\dots$). As the state we consider belongs to the $sl(2)$ subsector of the theory, the middle node Y-functions of two types, corresponding to two series of black nodes in figure \ref{Fig:figabjmysys}, are pairwise equal. We denote these Y-functions by $\Ym{a}$ or $Y_{a,0}$, so that we have $ Y_{{\fO}_{a}} = Y_{{\fo}_{a}} = \Ym{a}=Y_{a,0}$ for all $a$.

The TBA equations which describe the exact energy of the ground state in finite volume were proposed in \cite{Bombardelli:2009xz}, \cite{GFcp3}. In \cite{GFcp3} they were also extended via the contour deformation trick \cite{Bazhanov:1996aq} to excited states in the $sl(2)$ subsector, the resulting equations being:
\beqa
\label{YFE}
    \log Y_{\figF} &=&
    +K_{m-1}*\log\frac{1+1/Y_{\figb_{m}}}{1+Y_{\figp_{m}}}
    +2\mathcal{R}_{1m}^{(01)}*\log(1+Y_{\figm_m})
    +
    2\left[\log\frac{R_4^{(+)}}{{R_4^{(-)}}}\right]
    + i\pi
\\
\label{YfE}
    \log Y_{\figf }&=&
    -K_{m-1}*\log\frac{1+1/Y_{\figb_{m}}}{1+Y_{\figp_{m}}}
    -2\mathcal{B}_{1m}^{(01)}*\log(1+Y_{\figm_m})
    -
    2\left[\log\frac{B_4^{(+)}}{{B_4^{(-)}}}\right]
    -i\pi
\\
\label{YpE}
    \log Y_{{\figp}_{n}}&=&-K_{n-1,m-1}*\log(1+Y_{{\figp}_{m}})
    -K_{n-1}*\log\frac{1+Y_{\figF}}{1+1/Y_{\figf}}
\\
\nn
    &+&
    2\left({\cal R}^{(01)}_{nm}+{\cal
    B}^{(01)}_{n-2,m}\right)*\log(1+Y_{\figm_m})
\\
\nn
    &+&
    2\left[\sum\limits_{k=-\frac{n-1}{2}}^{\frac{n-1}{2}}
    \log\frac{R_4^{(+)} (u+ik)}{{R_4^{(-)} (u+ik)}} \right]
    +
    2\left[\sum\limits_{k=-\frac{n-3}{2}}^{\frac{n-3}{2}}
    \log\frac{B_4^{(+)} (u+ik)}{{B_4^{(-)} (u+ik)}} \right]
\\
\label{YbE}
    \log Y_{{\figb}_{n}}&=&K_{n-1,m-1}*\log(1+1/Y_{{\figb}_{m}})
    +K_{n-1}*\log\frac{1+Y_{\figF}}{1+1/Y_{\figf}}
\\
\label{YmE}
    \log Y_{{\figm}_{n}}&=&
    (L+K_4) \log\frac{x^{[-n]}}{x^{[+n]}}
    -
    {\cal B}_{n1}^{(10)}*\log(1+1/Y_{\figf})
    +{\cal R}_{n1}^{(10)}*\log(1+Y_{\figF})
\\
\nn
    &+&\left({\cal R}_{nm}^{(10)}+{\cal B}_{n,m-2}^{(10)}\right) * \log(1+Y_{\figp_m})
\\
\nn
    &+&
    \(2\tilde{\cal S}_{nm} - {\cal R}_{nm}^{(11)} + {\cal B}_{nm}^{(11)}\) * \log(1+Y_{\figm_m})
    + \left[ \sum_{k=-\frac{n-1}{2}}^{\frac{n-1}{2}} \log \Phi(u+ik)\right]+i\pi n
\eeqa
Here and throughout the paper we use notation that is given in Appendix A. We also assume summation over the repeated index $m$.

The exact Bethe roots $u_{4,j}=u_{\bar4,j}$ are fixed by the exact Bethe equations,
\beq
\label{Yism1}
        Y_{\fm_1}^\ph(u_{4,j}) = -1, \ \ \ \ j=1,2,\dots,K_4
\eeq
and in general the values of these Bethe roots may differ from those one gets from the ABA. In the case we are studying the roots remain at zero within our precision -- we have checked this numerically, verifying \eq{Yism1} with the use of equation \eq{YmE} which we analytically continued as in \cite{GKVKonishi}.

The energy of the state we are studying is written in terms of the Y-functions as
\beq
\label{Eabjm}
    E = E_{ABA}
         + \delta E,\ \ \ 
        \delta E=2\sum_{a=1}^{\infty}\int_{-\infty}^{\infty}
        \frac{du}{2\pi i}\frac{\d \epsilon_a^{\mir}(u)}{\d u}
    \log(1+Y^{\mir}_{\figm_a})
\eeq
where $\epsilon_a(u)$ is defined by \eq{epsilon}.

As our goal is to solve the TBA equations numerically, we will make to them several modifications, which are described below.

First, we substract from the original equations the equations which are satisfied by the asymptotic large $L$ solution of the Y-system \cite{GKV}, \cite{GFcp3}. Let us describe this solution for the state discussed in section 2 which belongs to the $sl(2)$ subsector and has $L=1$, $K_4=K_{\bar 4}=1$ with two Bethe roots $u_4=u_{\bar 4}=0$. We use bold font to denote the asymptotic Y- and T-functions. The main formulas are:\footnote{I am grateful to A. Cavaglia for letting me know of misprints in Eq. \eq{eqFF}.}
\beq\la{YbtoT}
    {\bf Y}_{\fp_a}=\frac{{\bf T}^+_{a,1}{\bf T}^-_{a,1}}{{\bf T}_{a+1,1}{\bf T}_{a-1,1}}-1\;\;,\;\;
    1/{\bf Y}_{\fb_s}=\frac{{\bf T}^+_{1,s}{\bf T}^-_{1,s}}{{\bf T}_{1,s+1}{\bf T}_{1,s-1}}-1
\eeq

\beq
\label{eqFF}
        {\bf Y}_{\ff} = \frac{{\bf T}_{2,3}{\bf T}_{2,1}}{{\bf T}_{3,2}{\bf T}_{1,2}},\ 
        {\bf Y}_{\fF} = \frac{{\bf T}_{1,2}{\bf T}_{1,0}}{{\bf T}_{2,1}{\bf T}_{0,1}}.\     
\eeq
All the ${\bf T}_{a,s}$ functions can be found from the generating functional ${\cal W}$. In particular

\beq
    {\cal W}=\sum_{s=0}^\infty {\bf T}_{1,s}(u+i\tfrac{1-s}{2})D^s\;\;,\;\;
    {\cal W}^{-1}=\sum_{a=0}^\infty (-1)^a {\bf T}_{a,1}(u+i\tfrac{1-a}{2})D^a\;.
\eeq
\beqa
\nn
        {\cal W}&=&
        \left(1-\(\frac{B_4^{(+)+}}{B_4^{(-)+}}\frac{R_4^{(+)-}}{R_4^{(-)-}}\)^2D\right)
        \left(1-\(\frac{R_4^{(+)-}}{R_4^{(-)-}}\)^2\,D\right)^{-1}\\
        &\times&
        \left(1-\(\frac{R_4^{(+)-}}{R_4^{(-)-}}\)^2\,D\right)^{-1}
        \left(1-D\right)
\eeqa
where $D=e^{-i\d_u}$. Finally,
\beq
        {\bf Y}_{\figm_a}(u)= \(\frac{x^{[-a]}}{x^{[+a]}}\)^L \Phi_a(u) {\bf T}_{a,1}(u).
\eeq

A remarkable property of these asymptotic Y-functions is that they satisfy a set of TBA-type equations. This is true not only for $\lambda\to0$, but also for finite values of $\lambda$ and $L$ (see e.g. \cite{Gromov:2011cx}, \cite{yasymp})\footnote{For the state we are studying we have also checked this numerically}. The equations satisfied by the asymptotic solution are the same as the original TBA equations given above, except that the terms in the r.h.s. which involve the $\Ym{n}$-functions should be omitted. After substracting these equations from the original ones and also slightly rewriting the kernels for more convenient numerics (similarly to \cite{GKVKonishi}) we get:

\beqa
\label{YFEn1}
    \log \frac{Y_{\figF}}{{\bf Y_{\fF}}} &=&
    +K_{m-1}*\log\frac{1+1/Y_{\figb_{m}}}{1+Y_{\figp_{m}}}
                                        \frac{1+{\bf Y}_{\figp_{m}}}{1+1/{\bf Y}_{\figb_{m}}}
    +2\mathcal{R}^{(0m)}*\log(1+Y_{\figm_m})
\\
%{\bf Y}
\label{YfEn1}
    \log \frac{Y_{\figf }}{{\bf Y}_{\figf }}&=&
    -K_{m-1}*\log\frac{1+1/Y_{\figb_{m}}}{1+Y_{\figp_{m}}}
                                        \frac{1+{\bf Y}_{\figp_{m}}}{1+1/{\bf Y}_{\figb_{m}}}
    -2\mathcal{B}^{(0m)}*\log(1+Y_{\figm_m})
\\
\label{YpEn1}
    \log \frac{Y_{{\figp}_{n}}}{{\bf Y}_{{\figp}_{n}}}
    &=&
    -K_{n-1,m-1}*\log\frac{(1+Y_{{\figp}_{m}})}{(1+{\bf Y}_{{\figp}_{m}})}
    -K_{n-1}*\log\frac{1+Y_{\figF}}{1+1/Y_{\figf}}\frac{1+1/{\bf Y}_{\figf}}{1+{\bf Y}_{\figF}}
\\
\nn
    &+&
    2{\cal M}_{nm}*\log(1+Y_{\figm_m})
\\
\label{YbEn1}
    \log \frac{Y_{{\figb}_{n}}}{{\bf Y}_{{\figb}_{n}}}
    &=&
    K_{n-1,m-1}*\log\frac{(1+1/Y_{{\figb}_{m}})}{(1+1/{\bf Y}_{{\figb}_{m}})}
    +K_{n-1}*\log\frac{1+Y_{\figF}}{1+1/Y_{\figf}}\frac{1+1/{\bf Y}_{\figf}}{1+{\bf Y}_{\figF}}
\\
\label{YmEn1}
    \log \frac{Y_{{\figm}_{n}}}{{\bf Y}_{{\figm}_{n}}}&=&
    -
    {\cal B}^{(n0)}*\log\frac{(1+1/Y_{\figf})}{(1+1/{\bf Y}_{\figf})}
    +{\cal R}^{(n0)}*\log\frac{(1+Y_{\figF})}{(1+{\bf Y}_{\figF})}
\\
\nn
    &+&{\cal N}_{nm}* \log\frac{(1+Y_{\figp_m})}{(1+{\bf Y}_{\figp_m})}
\\
\nn
    &+&
    \(2\tilde{\cal S}_{nm} - {\cal R}_{nm}^{(11)} + {\cal B}_{nm}^{(11)}\) * \log(1+Y_{\figm_m})
\eeqa
Note that in the Konishi case the function $\Yp{2}(u)$ had poles at the positions of the Bethe roots; in accordance with that, in our case it has a double pole at $u=0$. The asymptotic Y-function ${\bf Y}_{\figp_{2}}(u)$ also has a double pole at $u=0$, so the combination $\frac{1+{Y}_{\figp_{2}}(u)}{1+{\bf Y}_{\figp_{2}}(u)}$ which appears in the equations \eq{YFEn1}--\eq{YmEn1} has no singularities on the real axis.

In the next section we discuss a further simplification to these equations which leaves only a finite number of unknown functions.

\section{Truncating the TBA equations}
The truncation method which we describe in this section has been proposed in \cite{Gtrunc} and is a simpler version of the general treatment in \cite{Gromov:2011cx}. Unlike the method of \cite{Gromov:2011cx} it involves some approximations, and it also relies on certain analyticity assumptions for the Y-functions. We will not discuss these assumptions here, but we have confirmed numerically that in our case the resulting equations are consistent with the original TBA of \cite{GFcp3}.

The truncation is done as follows. First, remarkably, it is possible to eliminate all the $\Yb{n}$ functions from the TBA equations, replacing them by a single unknown function. The reason is that the Y-system equations for these functions are quite simple,
\beq
\label{ysysbos}
   {Y_{{\figb}_{n}}^+Y_{{\figb}_{n}}^-}
        =(1+Y_{{\figb}_{n+1}})(1+Y_{{\figb}_{n-1}}) \,\, , \,\, n\geq3
\eeq
and to solve them we can use the following ansatz for the corresponding T-functions (see \cite{Gromov:2008gj}, \cite{Gtrunc}, \cite{Gromov:2011cx})\footnote{see also \cite{Suzuki:2011dj}}:
\beq
\label{Tsf}
        T_{1,s}=s+K_s*f_R
\eeq
where $f_R$ is a new unknown function. The $\Yb{n}$ functions are obtained \cite{GKV}, \cite{GFcp3}, as usual, from
\beq
        {Y}_{\fb_n}=\(\frac{{T}^+_{1,n}{T}^-_{1,n}}{{ T}_{1,n+1}{T}_{1,n-1}}-1\)^{-1},
\eeq
and it can be shown that the infinite set of equations \eq{ysysbos} is indeed satisfied for any $f_R$ with suitable analyticity properties. This function is fixed by the TBA equations -- the infinite set of equations \eq{YbEn1} reduces to a single equation for $f_R$. That equation has the form
\beq
        \frac{1+\YF}{1+1/\Yf}=\frac{\(1+K_1^+*_{p.v.}f_R+f_R/2\)\(1+K_1^-*_{p.v.}f_R+f_R/2\)}
                                                                                {\(1+K_1^+*_{p.v.}f_R-f_R/2\)\(1+K_1^-*_{p.v.}f_R-f_R/2\)}
\eeq
where $*_{p.v.}$ denotes principal value integration. Solving for $f_R$ we get an expression suitable for numerical iterations\footnote{Note that $f_R$ is only nonzero for $-2h\leq u\leq 2h$.}:
\beq
\label{fReq}
        f_R = 2\frac{(a+1)(y+1)-\sqrt{4(a+1)^2y^2-b^2(y-1)^2}}{y-1}
\eeq
where $y=\frac{1+\YF}{1+1/\Yf}, \ a = \Re \(K_1^+*_{p.v.}f_R\),\ b = \Im \(K_1^+*_{p.v.}f_R\)$.

Note that the Y-system equation for $\Yb{2}$, which reads
\beq
        {Y_{{\figb}_{2}}^+Y_{{\figb}_{2}}^-}
        =\frac{(1+Y_{\figF})(1+Y_{{\figb}_{3}})}{1+1/Y_{\figf}},
\eeq
is not satisfied automatically for arbitrary $f_R$ (since it includes the fermionic functions $\Yf, \YF$), but will hold provided that the TBA equations are satisfied.

So far we haven't made any approximations in the TBA equations -- the truncation we have just described is exact. However, we cannot directly apply the same method in the upper wing of the Y-system to replace the functions $\Yp{n}, \Ym{n}$ by a finite number of unknown ones, as the Y-system equations for $\Yp{n}, \Ym{n}$ are more complicated than \eq{ysysbos}. But since the $\Ym{n}$ functions decay fast with $n$, within our precision it is enough to keep only the first 6 or 7 of them and set the other ones to zero. With this approximation we see that for $n>M$, where $M$ the number of $\Ym{n}$ functions that we retain, the pyramid Y-functions $\Yp{n}$ decouple from the rest of the Y-system and are governed by an equation very similar to \eq{ysysbos}:
\beq
\label{ysyspyr}
        {Y_{{\figp}_{n}}^+Y_{{\figp}_{n}}^-}
        =\frac{1}{(1+1/Y_{{\figp}_{n+1}})(1+1/Y_{{\figp}_{n-1}})} \,\, , \,\, n>M .
\eeq
This equation is solved by an ansatz analogous to \eq{Tsf}:
\beq
\label{Taf}
        T_{a,1}=a+K_{a-M+1}*f_U
\eeq
with ${Y}_{\fp_a}=\frac{{T}^+_{a,1}{T}^-_{a,1}}{{T}_{a+1,1}{T}_{a-1,1}}-1$. Here $f_U$ is another new unknown function, and it is fixed by an equation similar to \eq{fReq}:
\beq
\label{fUeq}
        f_U=a (4y-2)-2(\sqrt{(y-1)y (2a+M+1)^2-b^2}-(M+1)y+M)
\eeq
where $y=1+\Yp{M}, \ a = \Re \(K_1^+*_{p.v.}f_U\),\ b = \Im \(K_1^+*_{p.v.}f_U\)$.

Lastly, we need to rewrite the r.h.s. of the remaining TBA equations in terms of the new functions $f_R, f_U$. To do this, we will need the exact T-functions -- which are given by \eq{Tsf}, \eq{Taf} -- and the asymptotic T-functions, which are obtained from the same expressions when $f_R, f_U$ are replaced by ${{\bf f}_R},{{\bf f}_U}$:
\beq
\label{asympf}
                {{\bf f}_R}(u) = \frac{x^{\mir}(u-i0)-1/x^{\mir}(u+i0)}
                                                        {x_4^{\mir +}-1/x_4^{\mir -}},
                                                        \ \ 
                {{\bf f}_U}(u) = K_{M-1} (u,v) * \frac{x^{\mir}(v-i0)-1/x^{\mir}(v+i0)}
                                                        {\(1/x_4^{\mir +}\)-x_4^{\mir -}}.
\eeq
We list the remaining TBA equations below.

Equations for fermions:
\beqa
\label{YFF}
        \log \frac{Y_{\figF }}{{\bf Y}_{\figF }}
        &=&
        \log \frac{T_{1,2}}{{\bf T}_{1,2}}
    -K_{1}*\log\frac{T_{1,1}}{{\bf T}_{1,1}}
    -K_{M-1}*\log\frac{T_{M+1,1}}{{\bf T}_{M+1,1}}                                      
    +K_{M}*\log\frac{T_{M,1}}{{\bf T}_{M,1}}
\\ \nn
    &-&
    \sum_{m=2}^{M} K_{m-1}*\log\frac{1+Y_{\figp_{m}}}{1+{\bf Y}_{\figp_{m}}}
    +2\mathcal{R}^{(0m)}*\log(1+Y_{\figm_m})
\\
\label{YFf}   \nn 
    \log \frac{Y_{\figf }}{{\bf Y}_{\figf }}
    &=&
        -\log \frac{T_{1,2}}{{\bf T}_{1,2}}
    +K_{1}*\log\frac{T_{1,1}}{{\bf T}_{1,1}}
    +K_{M-1}*\log\frac{T_{M+1,1}}{{\bf T}_{M+1,1}}                                      
    -K_{M}*\log\frac{T_{M,1}}{{\bf T}_{M,1}}
\\ 
    &+&
    \sum_{m=2}^{M} K_{m-1}*\log\frac{1+Y_{\figp_{m}}}{1+{\bf Y}_{\figp_{m}}}
    -2\mathcal{B}^{(0m)}*\log(1+Y_{\figm_m}).
\eeqa
Equations for pyramids ($n=2,\dots,M$)
\beqa
\label{YpF}
\nn
    \log \frac{Y_{{\figp}_{n}}}{{\bf Y}_{{\figp}_{n}}}
    &=&
    -\delta_{n,M} \log\frac{T_{M+1,1}}{{\bf T}_{M+1,1}}
    -K_{n-1,M-1}*\log\frac{T_{M+1,1}}{{\bf T}_{M+1,1}}                                          
    +K_{n-1,M  }*\log\frac{T_{M,1  }}{{\bf T}_{M,1  }}
\\ 
    &-&
    \sum_{m=2}^{M} K_{n-1,m-1}*\log\frac{1+Y_{\figp_{m}}}{1+{\bf Y}_{\figp_{m}}}
    -K_{n-1}*
    \log\frac{1+Y_{\figF}}{1+1/Y_{\figf}}\frac{1+1/{\bf Y}_{\figf}}{1+{\bf Y}_{\figF}}
\\
\nn
    &+&
    2{\cal M}_{nm}*\log(1+Y_{\figm_m}).
\eeqa
Equations for middle nodes ($n=1,2,\dots,M$):
\beqa
\label{YmF}
    \log \frac{Y_{{\figm}_{n}}}{{\bf Y}_{{\figm}_{n}}}
    &=&
    \(2\tilde{\cal S}_{nm} - {\cal R}_{nm}^{(11)} + {\cal B}_{nm}^{(11)}\) *  
     \log(1+Y_{\figm_m})
\\
\nn
    &-&{\cal B}^{(n0)}*\log\frac{(1+1/Y_{\figf})}{(1+1/{\bf Y}_{\figf})}
    +{\cal R}^{(n0)}*\log\frac{(1+Y_{\figF})}{(1+{\bf Y}_{\figF})}\\
\nn
    &+&
    \mathcal{R}^{(n0)}\cutclock
    \left(K_{M-1}*\log\frac{T_{M+1,1}}{{\bf T}_{M+1,1}}                                     
    -K_{M  }*\log\frac{T_{M,1  }}{{\bf T}_{M,1  }}
    +\sum_{m=2}^{M} K_{m-1}*\log\frac{1+Y_{\figp_{m}}}{1+{\bf Y}_{\figp_{m}}}
    \right)
\\
\nn
        &+&
        K^{\neq}_{n-1,M-1}*\log\frac{T_{M+1,1}}{{\bf T}_{M+1,1}}                                        
   -K^{\neq}_{n-1,M  }*\log\frac{T_{M,1  }}{{\bf T}_{M,1  }}
   + \sum_{m=2}^{M} K^{\neq}_{n-1,m-1}*\log\frac{1+Y_{\figp_{m}}}{1+{\bf Y}_{\figp_{m}}} .
\eeqa
As a result, the equations that we are solving numerically are \eq{fReq}, \eq{fUeq} and \eq{YFF}--\eq{YmF}.

\section{Numerical results}

%Our numerical strategy is similar to \cite{GKVKonishi}. We solved the TBA equations \eq{fReq}, \eq{fUeq}, \eq{YFF}--\eq{YmF} iteratively using a standard relaxation method\footnote{Sometimes we had to iterate the equation for $\Ym{1}$ separately from the others to make the process converge.}: plugging the functions $Y_{a,s}^{(N)}$ that we obtained on step $N$ into the r.h.s. of the TBA equations we read off from the l.h.s. the prediction $\tilde{Y}_{a,s}^{(N+1)}$ for new Y-functions. We then used a weighted sum of this prediction and the previous result\footnote{In other words, we are using a standard relaxation method, discussed recently in the TBA context in e.g. \cite{}} as the Y-functions on step $N+1$, i.e. $Y_{a,s}^{(N+1)}=\alpha\tilde{Y}_{a,s}^{(N+1)}+(1-\alpha)Y_{a,s}^{(N)}$ where usually the weight $\alpha$ is taken to be $0.25$.

Our numerical strategy is similar to \cite{GKVKonishi}, and we are solving the TBA equations \eq{fReq}, \eq{fUeq}, \eq{YFF}--\eq{YmF} by iterations\footnote{Sometimes we also had to iterate the equation for $\Ym{1}$ separately from the others to make the process converge.}. We truncated the number of middle node Y-functions to $M=6$ or 7, and used cutoffs on the $u$ axis to make finite the range of integration in the convolutions. The range of coupling that we study is $0\leq h(\lambda)\leq1$ (since the TBA equations are written in terms of $h(\lambda)$, we obtain the anomalous dimension as a function of $h(\lambda)$).

\TABLE[ht]{\label{Table:ewrap}
\caption{Numerical values of the correction to the Bethe ansatz result for the energy, with uncertainty in the last digit.}
\begin{tabular}{|c|r||c|r|}
  \hline
  $h(\lambda)$ &$\delta E(\lambda)$ & $h(\lambda)$ &$\delta E(\lambda)$
  \\ \hline
 0.00 & 0.0000 & 0.55 & 0.0703 \\
 0.10 & 0.0005 & 0.60 & 0.069 \\
 0.15 & 0.0023 & 0.65 & 0.059 \\
 0.20 & 0.0063 & 0.70 & 0.041 \\
 0.25 & 0.0129 & 0.75 & 0.014 \\
 0.30 & 0.0221 & 0.80 & -0.025 \\
 0.35 & 0.0332 & 0.85 & -0.072 \\
 0.40 & 0.0451 & 0.90 & -0.126 \\
 0.45 & 0.0566 & 0.95 & -0.188 \\
 0.50 & 0.0655 & 1.00 & -0.254 \\
\hline
\end{tabular}
}

Our main result is the correction $\delta E$ to the ABA value for the energy (see \eq{eplusde}) and it is shown in Fig. \ref{Fig:ewrap}. The numerical values of this correction are also listed in table \ref{Table:ewrap}. We estimated their absolute precision as about $\pm 10^{-4}$ for $h(\lambda)\leq 0.55$ and $\pm 10^{-3}$ for $h(\lambda)> 0.55$. In Figure \ref{Fig:ebae} we plot the full energy, including the ABA part. As expected, at weak coupling our numerical results are completely consistent with the leading wrapping correction obtained analytically in \cite{GKV}, \cite{MSS2} (this is also seen in Figure \ref{Fig:ewrap}). As the coupling is increased, our results gradually deviate more and more from that prediction.

\FIGURE[ht]{\label{Fig:ewrap}

    \begin{tabular}{cc}
    \includegraphics{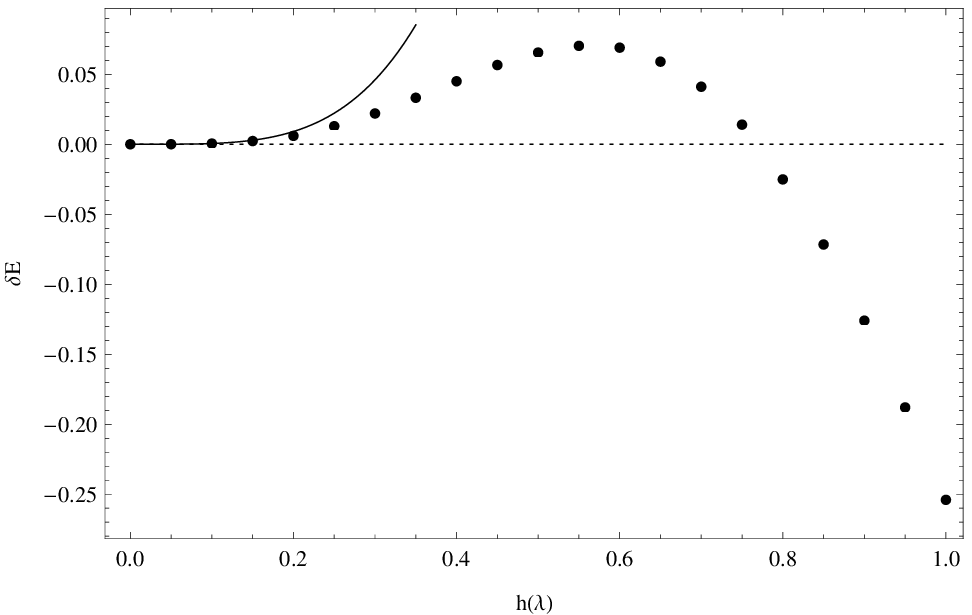}
    \end{tabular}
    \caption{The correction $\delta E$ to the Bethe ansatz result for the energy, as a function of $h(\lambda)$, is shown by black dots. The solid line is the first wrapping correction given by \eq{1wrap}.}
}

The behaviour of middle-node Y-functions $\Ym{n}$ exhibits several interesting features. While in the AdS$_5$/CFT$_4$ Konishi case all these functions take positive values, here they have signs alternating with $n$, i.e. $\Ym{1}<0,\ \Ym{2}>0, \ \Ym{3}<0$ etc. In Figure \ref{Fig:allym} (left) we plot their absolute value for $h(\lambda)=0.9$. Since these Y-functions appear as $\log(1+\Ym{n})$ in the expression for the energy and in the TBA equations, their negative values beyond $-1$ would give rise to singularities. We discovered that as the coupling is increased, the middle-node Y-function $\Ym{1}$ indeed approaches the critical value $-1$ for $u$ close to zero (while other Y-functions remain far from this dangerous value). That is clearly seen from Figure \ref{Fig:ym1allg} (left), and at $h(\lambda)=1$ we have $\Ym{1}(0)\approx-0.993$ which is already very close to the singularity. This means that proceeding further in the coupling would probably require obtaining $\Ym{1}$ with a high precision.

\FIGURE[ht]{\label{Fig:ebae}

    \begin{tabular}{cc}
   \includegraphics{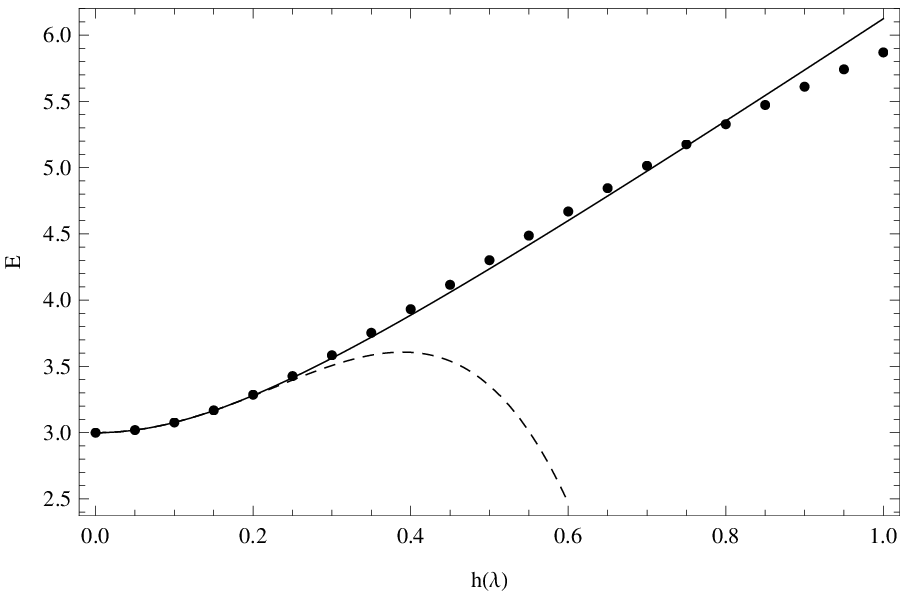}\\
    \end{tabular}
    \caption{The full scaling dimension $E_{ABA}+\delta E$. Dots: numerics, solid line: ABA, dashed line: ABA + 1st wrapping as given by \eq{aba1stwrap}}
}

The corresponding {\it asymptotic} Y-function ${\bf Y}_{\figm_1}$ already starts to take values beyond $-1$ at $h(\lambda)\approx0.6$. However, that does not lead to any singularity since $\log(1+{\bf Y}_{\figm_1})$ never appears in the TBA equations. In Figure \ref{Fig:allym} (right) we plot for comparison the exact and the asymptotic Y-function.

\FIGURE[ht]
{\label{Fig:allym}

    \begin{tabular}{cc}
    \includegraphics[scale=0.7]{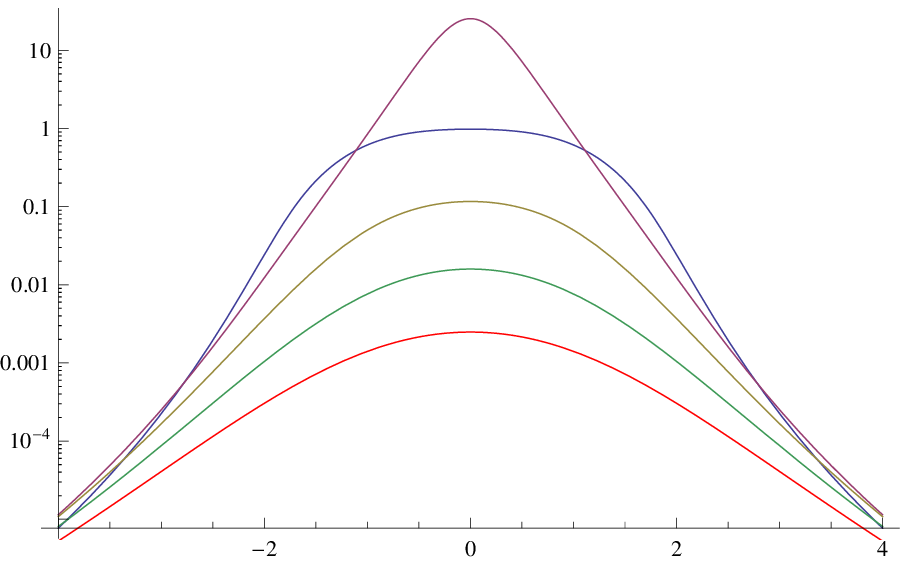}&\includegraphics[scale=0.7]{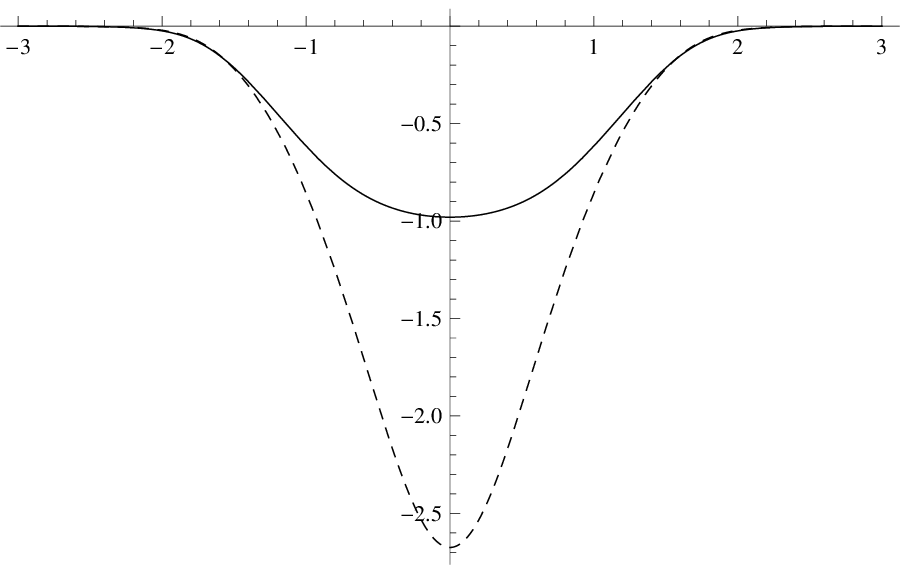}\\
    \end{tabular}
    \caption{{\bf Left}: The middle-node Y-functions $Y_{n,0} (u)$ for $h(\lambda)=0.9$. The figure shows plots of minus $Y_{1,0}$ (blue), $Y_{2,0}$ (purple), minus $Y_{3,0}$ (yellow), $Y_{4,0}$ (green) and minus $Y_{5,0}$ (red).  {\bf Right}: the exact Y-function $Y_{1,0}$ (solid line) and its asymptotic counterpart ${\bf Y}_{1,0}$ (dashed line) for $h(\lambda)=0.9$.}
    }

It is possible that if the coupling is increased further the function $\Ym{1}$ will cross the critical value $-1$. Then the TBA equations will probably require some modification such as extra driving terms, and it would be very interesting to understand whether this indeed happens for the state we are studying. For other models similar issues have been explored in \cite{Bazhanov:1996aq}, while in the AdS/CFT case the possibility of such singularities arising was discussed in e.g. \cite{GKVKonishi}, \cite{Arutyunov:2009ax},  \cite{Arutyunov:2010gu}.

The second middle-node Y-function $\Ym{2}$ also shows unusual behaviour, rapidly increasing in magnitude as the coupling is being increased. This is shown in Figure \ref{Fig:ym1allg} (right).

\FIGURE[ht]
{\label{Fig:ym1allg}

    \begin{tabular}{cc}
    \includegraphics[scale=0.75]{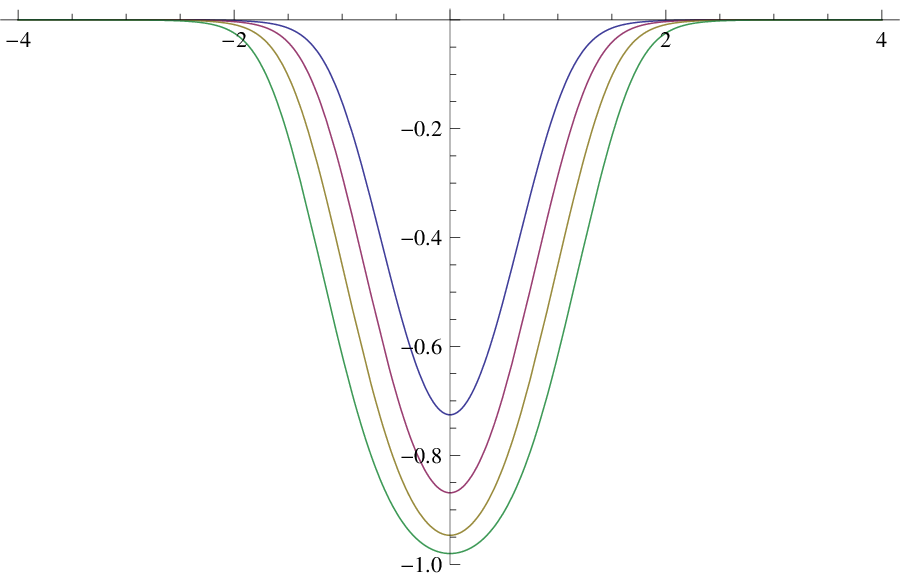}&\includegraphics[scale=0.75]{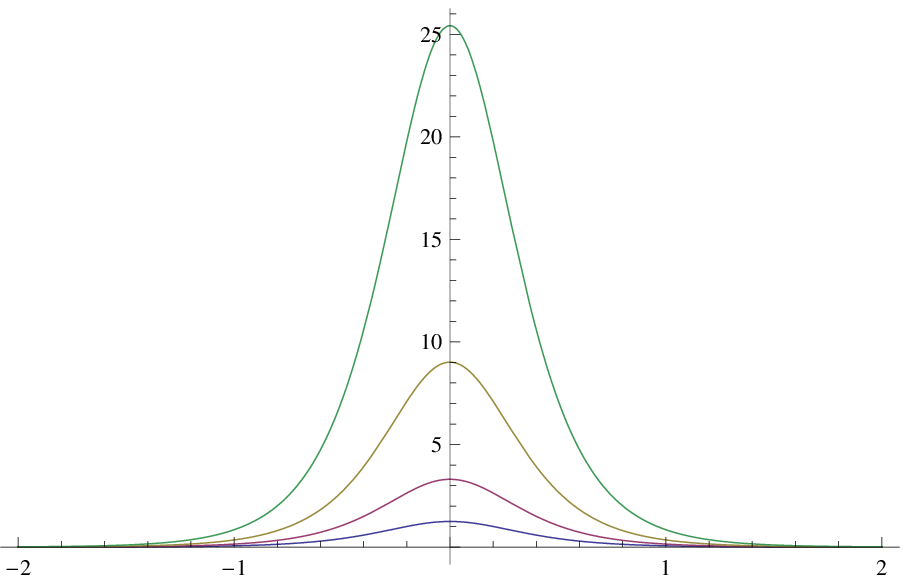}\\
    \end{tabular}
    \caption{The middle-node Y-functions $Y_{1,0}(u)$ (left) and $Y_{2,0}(u)$ (right) for several values of the coupling:  $h(\lambda)=0.6$ (blue), 0.7 (purple), 0.8 (yellow) and 0.9 (green).}
    }

Lastly, in Figure \ref{Fig:rightf90} we show the plots of the new functions $f_R, f_U$ which parametrize the solution of the Y-system in the right wing and in the upper wing, respectively.

\FIGURE[ht]
{\label{Fig:rightf90}

    \begin{tabular}{cc}
    \includegraphics[scale=0.7]{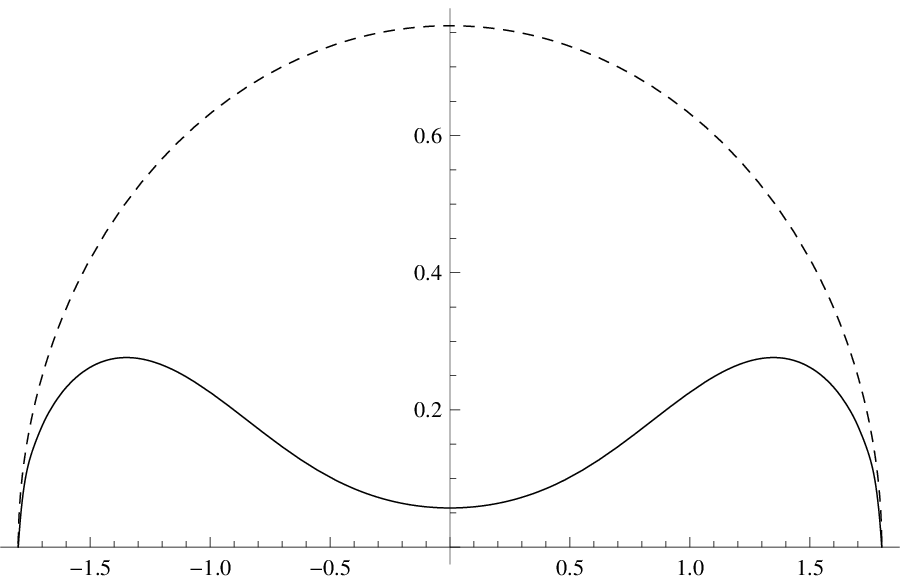}&\includegraphics[scale=0.7]{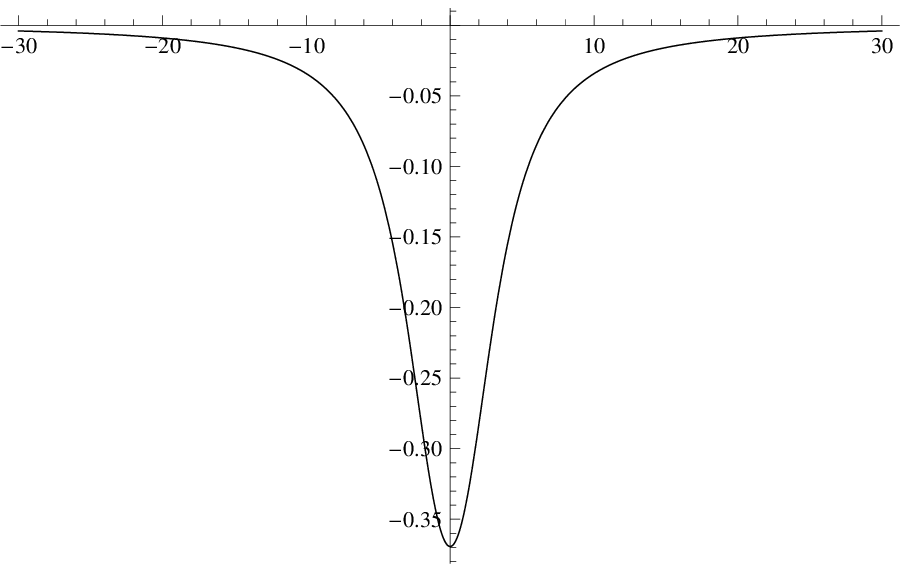}\\
    \end{tabular}
    \caption{The functions $f_R(u)$ (left) and $f_U(u)$ (right) for $h(\lambda)=0.9$ and $M=7$, shown by solid lines. The dashed line in the left figure shows the asymptotic ${\bf f}_R$ function \eq{asympf}, while $f_U$ almost coincides with its asymptotic expression ${\bf f}_U$.}
    }

The total range of the coupling we have investigated, $0\leq h(\lambda)\leq1$, is four times greater than the convergence radius $|h(\lambda)|\leq0.25$ of the weak-coupling expansion of the ABA result \eq{eaba}, which suggests that we are exploring the intermediate coupling regime. We were not able to make a consistent prediction for the strong coupling expansion coefficients, but we hope this could be done in the future by going to greater values of the coupling. Increasing the coupling poses a challenge because beyond the value $h(\lambda)=1$ the iterative procedure we use for solving the TBA equations converges too slowly; hopefully this problem may be overcome by improving the numerical algorithm (e.g. using Newton's method).

At strong coupling the leading term in the exact energy should be proportional to $\lambda^{1/4}$ \cite{Gubser:1998bc} which in our case is equivalent to $\sqrt{h(\lambda)}$ (while string theory predictions for subleading terms are not available as of now). However, the strong-coupling behavior of the ABA result \eq{eaba} is completely different
\beq
\label{abastrong}
	E_{ABA}=2\sqrt{2\lambda}+O(1)=4h(\lambda)+O(1),
\eeq
suggesting that the leading asymptotics predicted by the ABA should cancel against\footnote{In contrast, in the AdS$_5$/CFT$_4$ Konishi case
\cite{GKVKonishi, Frolov:2010wt, KonOther, Gromov:2011de}
%\cite{GKVKonishi}, \cite{Frolov:2010wt}, \cite{KonOther}, \cite{Gromov:2011de}
the ABA result already has the correct $\sim \lambda^{1/4}$ asymptotics at strong coupling. For Konishi the TBA correction $\delta E$ then compensates another correction which arises because the exact Bethe roots are displaced from their ABA values, and as a result the exact energy has the same asymptotics.} the TBA correction $\delta E$. It is possible that we can already see this starting to happen for $h(\lambda)\geq0.8$ when the exact energy becomes smaller than the ABA prediction. In Figure \ref{Fig:ebaeloglog} we show a log-log plot of the full energy, and the ABA result asymptotes to a straight line, consistently with \eq{abastrong}, while one may expect that the exact energy will asymptote to a straight line with a different slope\footnote{not necessarily the dashed line shown in Figure \ref{Fig:ebaeloglog} }. 

\FIGURE[ht]
{\label{Fig:ebaeloglog}

    \begin{tabular}{cc}
    \includegraphics[scale=1]{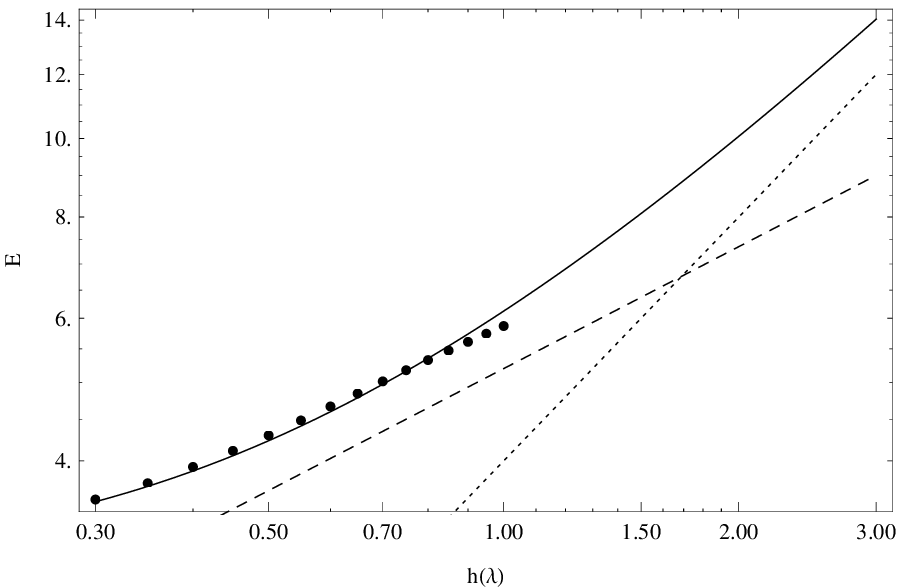}\\
    \end{tabular}
    \caption{A log-log plot of the full conformal dimension $E$ vs. $h(\lambda)$. We show the ABA prediction (solid curve), its asymptote (dotted line defined by $E=4h(\lambda)$), our numerical data (black dots) and the expected slope of the result at strong coupling (dashed line defined by $E=\const\times\sqrt{h(\lambda)}$).  }
    }

\section{Conclusions}

In this paper we have applied the Thermodynamic Bethe ansatz approach to compute the exact anomalous dimension of a short operator in AdS$_4$/CFT$_3$, for the first time solving numerically the TBA equations for this theory. We have explored the values of the coupling $0\leq h(\lambda)\leq 1$ and at weak coupling our results are consistent with known predictions. We also found that as the coupling is being increased one of the Y-functions approaches the critical value $-1$. It would be very interesting to go to higher values of the coupling, and achieving that may be possible by improving the numerical strategy or perhaps by applying the very recent approach of \cite{Gromov:2011cx}. Hopefully a comparison with string theory calculations could be made eventually at strong coupling, providing  a deep test of the integrable structures in AdS$_4$/CFT$_3$ correspondence.

\section*{Acknowledgements}
I thank M. Beccaria, N. Gromov, A. Tseytlin and A. Zayakin for many helpful comments and discussions. I am especially grateful to Nikolay Gromov for sharing some of his Mathematica code and for his guidance during the course of this project. This work was partially supported by a grant of the Dynasty Foundation and by grants RFBR-12-02-00351-a, PICS-12-02-91052. I am also grateful to the organizers of IGST 2011 at Perimeter Institute (Waterloo, Canada) for hospitality during this program.

\section{Appendix A: Notation}

The R and B functions are
\be
R_a^{(\pm)} =  \prod_{j=1}^{K_{a}}\left[x(u)-x^{\mp}_{a,j}\right], \qquad
B_a^{(\pm)} =  \prod_{j=1}^{K_{a}}\left[\frac{1}{x(u)}-x^{\mp}_{a,j}\right],
\ee
where for the state we consider in this paper all $K_a$ are zero except $K_4 = K_{\bar4}=1$. The scalar factor $\Phi$ is defined as
\beq
\la{defPhi}
        \Phi(u) = \frac{B_4^{(+)+}R_{ 4}^{(-)-}}{B_4^{(-)-}R_{4}^{(+)+}}
        \(\prod_{j=1}^{K_4}\frac{x^+_{4,j}}{x^-_{4,j}}\)\prod_{j=1}^{K_4}\sigma^2(u,u_{4,j}), \ \ 
        \Phi_a(u)=\prod^{\frac{a-1}{2}}_{k=-\frac{a-1}{2}}\Phi(u+ik).
\eeq
The kernels in TBA equations are:
    \beqa
        K_n (u, v) &\equiv& \frac{1}{2\pi i} \frac{\d}{\d v} \ln\frac{u-v+in/2}{u-v-in/2}
        = \frac{2n/\pi}{n^2+4(u-v)^2},
    \\
        K_{n,m}(u,v) &\equiv&
        \sum_{j=-\frac{m-1}{2}}^{\frac{m-1}{2}}\sum_{k=-\frac{n-1}{2}}^{\frac{n-1}{2}}
        K_{2j+2k+2}(u,v),
      \\
        K_{n,m}^{\neq}(u,v) &\equiv&
        \sum_{j=-\frac{m-1}{2}}^{\frac{m-1}{2}}\sum_{k=-\frac{n-1}{2}}^{\frac{n-1}{2}}
        K_{2j+2k+1}(u,v),
\eeqa
where we assume $K_{0,n}=0, \ K_{0,n}^{\neq}=0$. Also,
\beqa
    \\
        {\cal S}_{nm}(u,v) &\equiv&
        \frac{1}{2\pi i}\frac{\d}{\d v}
        \log \sigma_{BES}(x^{[+n]}(u), x^{[-n]}(u), x^{[+m]}(v), x^{[-m]}(v)),
    \\
        \tilde {\cal S}_{nm}(u,v) &\equiv& {\cal S}_{nm}(u,v) +\frac{ni}{2}{\cal P}^{(m)}(v),
    \\
        \mathcal{ B}^{(ab)}_{nm}(u,v) &\equiv&
        \sum_{j=-\frac{n-1}{2}}^{\frac{n-1}{2}}\sum_{k=-\frac{m-1}{2}}^{\frac{m-1}{2}}
        \frac{1}{2\pi i}\frac{\d}{\d v} \log
        \frac{b(u+ia/2+ij,v-ib/2+ik)}{b(u-ia/2+ij,v+ib/2+ik)}
    \\
        \mathcal{ R}^{(ab)}_{nm}(u,v) &\equiv&
        \sum_{j=-\frac{n-1}{2}}^{\frac{n-1}{2}}\sum_{k=-\frac{m-1}{2}}^{\frac{m-1}{2}}
        \frac{1}{2\pi i}\frac{\d}{\d v} \log
        \frac{r(u+ia/2+ij,v-ib/2+ik)}{r(u-ia/2+ij,v+ib/2+ik)},
    \\
    {\cal R}^{(nm)}(u,v)&\equiv&
        \frac{\d_v}{2\pi i}\log \frac{x_u^{[+n]}-x_v^{[-m]}}{x_u^{[-n]}-x_v^{[+m]}}-\frac{1}{2i}{\cal
        P}^{(m)}(v) ,
        \\
        {\cal B}^{(nm)}(u,v)&\equiv&
        \frac{\d_v}{2\pi i}\log \frac{{1/x}_u^{[+n]}-x_v^{[-m]}}
        {{1/x}_u^{[-n]}-x_v^{[+m]}}-
        \frac{1}{2i}{\cal P}^{(m)}(v)
        \\
                {\cal M}_{nm}&\equiv& K_{n-1}\cutclock {\cal R}^{(0m)}+K^{\neq}_{n-1,m-1} \,,
        \\
                {\cal N}_{nm}&\equiv&{\cal R}^{(n0)}\cutclock K_{m-1}+K^{\neq}_{n-1,m-1} \,,
  \eeqa
where
    \be
        b (u,v) = \frac{1/x^{\mir}(u) - x^{\mir}(v)}{\sqrt{x^{\mir}(v)}}, \ \
        r (u,v) = \frac{x^{\mir}(u) - x^{\mir}(v)}{\sqrt{x^{\mir}(v)}},
    \ee
    and
    \beq
    {\cal P}^{(a)}(v)=-\frac{1}{2\pi}\d_v\log\frac{x^{\mir}(v+ia/2)}{x^{\mir}(v-ia/2)}\;.
    \eeq

The convolutions in integral equations are over the second variable, and $*$ denotes standard convolution along the real axis: $K(u,v) * f(v) \equiv \int^{+\infty}_{-\infty} dv K(u,v) f(v)$.

The symbol $\cutclock$ denotes convolution along with analytical continuation across the cut $u\in (-\infty,-2h)\cup(2h,+\infty)$ (see \cite{GFcp3}, \cite{GKVKonishi}). E.g.
\beq
{\cal R}^{(n0)}\cut\log(1+Y_{\figF})=
\int_{-2h}^{2h} dv \left[ {\cal R}^{(n0)}\log(1+Y_{\figF})-{\cal B}^{(n0)}\log(1+1/Y_{\figf})
\right]
.
\eeq
One should also be careful about which branch of $x(u)$ to use in various places in TBA equations. The mirror branch is used when $u$ is the free variable or the variable that is being integrated over. Otherwise the physical branch is used.

    \end{document}